\documentclass[twocolumn,prc,aps,floatfix]{revtex4}
\usepackage{dcolumn}
\usepackage[dvips]{graphicx}
\begin{document}
\newcommand{\be}{\begin{eqnarray}}
\newcommand{\ee}{\end{eqnarray}}
\title{Enhanced violation  of the Lorentz invariance and  Einstein's equivalence principle  in nuclei and atoms}
\author{V.V. Flambaum}
\affiliation{
 School of Physics, University of New South Wales, Sydney 
2052, Australia and Johannes Gutenberg University, 55099 Mainz, Germany
}
\date{\today}
\begin{abstract}
Local Lorentz Invariance violating (LLIV) and Einstein equivalence principle violating (EEPV)  effects in atomic experiments are discussed. The EEPV effects are strongly enhanced in the narrow 7.8 eV transition in  $^{229}_{90}$Th nucleus.  Nuclear LLIV tensors describing anisotropy in the maximal attainable speed  for massive particles (analog of the Michelson-Morley experiment for light)
are expressed in terms of the experimental values of nuclear quadrupole moments. 
 Calculations for  nuclei of experimental interest 
  $^{133}_{55}$Cs, $^{85}_{37}$Rb, $^{87}_{37}$Rb, $^{201}_{80}$Hg, $^{131}_{54}$Xe and  $^{21}_{10}$Ne have been performed. The results for $^{21}_{10}$Ne are used to improve the limits on the proton LLIV interaction constants by 4 orders of magnitude.
\end{abstract}
\maketitle
Lorentz invariance and Einstein equivalence principle are in the foundation of the general relativity theory.  Observation of LLIV and EEPV may pave the way to a new, more general  theory  (see e.g.  \cite{KosRus11,KosteleckyPottingPRD1995,Horava,Pospelov}).   Since  a typical energy scale of LLIV and EEPV is assumed to be a very large  Plank scale of the quantum gravity,  the low-energy manifestations of LLIV and EEPV are expected to be extremely small. The aim of this paper is to identify enhanced effects of LLIV and EEPV which may be easier to detect, and provide a new interpretation of existing experiments which leads to significantly improved limits on certain LLIV interaction constants. 

A special role in the foundation of the relativity theory was played by the Michelson-Morley experiment searching for the anisotropy in the speed of light. Similar anisotropy in the maximal attainable speed for massive particles has been constrained for nucleons by NMR experiments ( see e.g.  \cite{Romalis,Lamoreaux,Chupp,Prestage,Cs}) and for electrons using optical atomic transitions \cite{Hohensee2013c,Pruttivarasin2014}.
Corresponding interaction in the non-relativistic limit  may be described by the anisotropic kinetic energy term    $c_{ab}(p^2 \delta_{ab} -3 p_a p_b)/6m$ where $p$ and $m$ are the particle momentum and mass, and  $c_{ab}$ are the constants characterising the magnitude of LLIV. Here we use notations of the works \cite{KosRus11,KosLan99,Bluhm,Colladay1997,Colladay1998} coming under the name of the Standard Model Extension (SME).  EEPV effects within SME  originate from the dependence of the effective mass on the gravitational potential $U_g$ which in the non-relativistic limit is described by the $U_g$-dependent correction to the kinetic energy
\begin{equation}\label{eq1}
\delta H=c_{00} \frac{2U_g}{3c^{2}} \frac{p^2}{2m},
\end{equation}
where  $c$ is the speed of light. Purely gravitational limit on the proton  EEPV interaction constant $c_{00}=(0.20 \pm 30) \cdot 10^{-6}$ has been obtained from the nuclear bound kinetic energy \cite{Hohensee2013}. The limit on the electron EEPV constant $c_{00} =(0.14 \pm 0.28) \cdot 10^{-7}$ was obtained from the Dy atom spectroscopy  \cite{Hohensee2013c}. Relativistic effects in SME have been discussed in Ref. \cite{Altschul}. 

\section{Equivalence principle violation} 
The gravitational potential  $U_g$ depends on the distance from Earth to Sun and oscillates with the period of one year due to the ellipticity of the Earth orbit. The amplitude of the oscillations is  $\Delta U_g/c^2=1.7 \cdot 10^{-10}$. Corresponding shift  in a transition frequency  $ \delta \omega$ due to the perturbation (\ref{eq1}) allows one  to extract the limit on $c_{00}$.
To measure $c_{00}$ we should consider transitions where the kinetic energy $\mathbf{p}^2/2m$ changes. Using the virial theorem  for the Coulomb interaction we obtain the difference of kinetic energies between the  final  and initial  states  
\begin{equation}\label{eq2}
-(<\mathbf{p}^2/2m_e>_{f} - <\mathbf{p}^2/2m_e>_{i})=  \hbar \omega \,.
\end{equation}
However, the relativistic corrections (which are not small in heavy atoms,  $\sim Z^2 \alpha^2 \sim 1$) violate the non-relativistic relation in  Eq. (\ref{eq2}), i.e. the effect does not vanish for the transition frequency  $ \omega $ tends to zero. This way we get a relative enhancement of the effect   $ \delta \omega /\omega $ when $ \omega $ is very small 
\cite{Hohensee2013c}. This is the case for  the experiment Ref. \cite{Hohensee2013c} where the transition frequency $ \omega + \delta \omega $ between nearly degenerate levels of Dy atom has been measured as a function of the Sun gravitational potential and the limit on $c_{00}$ in the electron sector  has been extracted. 

  To have larger values of the EEPV and  LLIV effects  we proposed to use  $4f-6s$ transition in Yb$+$ ion  \cite{Dzuba2015} and in heavy highly charge ions \cite{Dzuba2016} near the the crossing points of electron  energy levels found in Refs. \cite{Flambaum2009,Berengut2010,Berengut2012a,Berengut2012b,Berengut2012c,Safronova2014,Widenberger2015}  where electrons have large kinetic energy.
  
In the present work we show that a significantly  larger enhancement of the  EEPV effect, $\sim 10^5$ times,  may exist near the nuclear energy level "crossing", especially in  the narrow 7.8 eV  transition between the ground state and first excited state in  $^{229}$Th nucleus. The reason for the enhancement is that the nuclear (MeV) scale of the kinetic energy difference $<\mathbf{p}^2/2m>_{f} - <\mathbf{p}^2/2m>_{i}$ is very large in comparison with the atomic (eV) scale while the transition energy 7.8(5)  eV  \cite{Beck}  happens to be on the atomic scale and may be investigated using  laser spectroscopy. This narrow transition with the width $\sim 10^{-3}-10^{-4}$ Hz \cite{ Tkalya2015,th2} was suggested as an extremely high precision nuclear clock \cite{th4,Campbell,Rellergert,Peik2015}, where possible effects of the space-time variation of the fundamental constants are enhanced up to 5  orders of magnitude  \cite{th1} (see also  \cite{He,Hayes,Wiringa,Litvinova,Dmitriev}), and as a nuclear laser  \cite{Tkalya-11}
 

 Let us start from a simple analytical estimate of EEPV effect. The ground state of  $^{229}$Th nucleus is $J^P[Nn_z\Lambda]=5/2^+[633]$,
i.e. the deformed oscillator quantum numbers are $N=6$, $n_z=3$, the
 projection of the valence neutron orbital angular momentum on the
nuclear symmetry axis (internal  z-axis)
 is $\Lambda=3$, the spin projection $\Sigma=-1/2$, and the
 total angular momentum and the total 
 angular momentum projection are $J=\Omega=\Lambda+\Sigma=5/2$.
The 7.8 eV excited state is $J^P[Nn_z\Lambda]=3/2^+[631]$, i.e.
it has 
$\Lambda=1$, $\Sigma=1/2$ and  $J=\Omega=3/2$.

To estimate the effect of the perturbation (\ref{eq1}) we should find the difference of the kinetic energies in the excited and ground  states. The energy of both states
may be described by an equation
\begin{equation}\label{eq3}
E=<\frac{p^2}{2m_n}> +<U> + C \Lambda \Sigma ,
\end{equation}
where $m_n$ is the neutron mass and  the spin-orbit interaction constant in Th nucleus is $C= - 0.85$ MeV \cite{BM}.
In the simplest single-particle model  the difference of the spin-orbit energies between the excited ( $\Lambda=1$, $\Sigma=1/2$) and the ground ($\Lambda=3$, $\Sigma=-1/2$) states is $2 C$.
The many-body corrections reduce this difference to $1.2 C$ \cite{Wiringa}.

In the oscillator potential the kinetic and  potential energies are equal, $<\frac{p^2}{2m}> =<U>$.
This gives us an estimate for the  the difference of the neutron kinetic energies between  the excited and ground states 
\begin{eqnarray}\label{eq4}
\nonumber
<\mathbf{p}^2/2m_n>_{exc} - <\mathbf{p}^2/2m_n>_{gr}= \\
 0.5 (\hbar \omega -1.2 C)= 0.5\, MeV.
\end{eqnarray}
This simple analytical estimate shows that the Lorentz invariance and Einstein equivalence principle violating effects Eq.(\ref{eq1}) in $^{229}$Th nucleus may be  $10^5$ times larger than in atoms.

 This estimate agrees with the sophisticated many-body numerical calculations of the difference of the kinetic energies in the  excited and ground  states performed in Ref.  \cite{Litvinova}. 
The result of the most complete calculation performed using the Hartree-Fock-Bogolubov method  (with the pairing included) and the nucleon-nucleon interaction which gives  the calculated energy difference between the excited and ground  states ($\hbar \omega =$- 46 KeV)  closest to the experimental energy difference 7.8 eV, is  
\begin{equation}\label{eq5}
<\mathbf{p}^2/2m_n>_{exc} - <\mathbf{p}^2/2m_n>_{gr}= 0.954 \,  MeV.
\end{equation}
This is the neutron contribution. There is also the proton contribution:
\begin{equation}\label{eq6}
<\mathbf{p}^2/2m_p>_{exc} - <\mathbf{p}^2/2m_p>_{gr}= 0.233 \,  MeV.
\end{equation}
Thus both neutron and proton interaction constants $c_{00}$  may be measured in 7.8 eV Th transition. 
We see again that the difference of the kinetic energies is significantly larger than the transition energy , i.e. we have the enhancement  of the effect of EEPV effect in  Eq.  (\ref{eq1}).

 
  A similar value 
  $<\mathbf{p}^2/2m_n>_{exc} - <\mathbf{p}^2/2m_n>_{gr} \sim $ 0.1 - 1 MeV
  is expected in $^{235}$U 73 eV transition. The laser spectroscopy  for this transition will be available soon \cite{Ye}.  Another possibility is M\"ossbauer transitions. However, the frequencies in these transitions are larger, so the relative effects will be smaller. The accuracy of the measurements in the M\"ossbauer spectroscopy is also not as high as in the laser spectroscopy.
  
  {\bf Spacial tensor  LLIV interaction in spherical nuclei: the Schmidt model and semi-empirical approach}. Now we consider effects  of  the tensor LLIV interaction 
   $c_{ab}(p^2 \delta_{ab} -3 p_a p_b)/6m$. The limits on  $c_{ab}$ in the electron sector have recently been obtained in Ref. \cite{Hohensee2013c} and significantly improved in Ref. \cite{Pruttivarasin2014}. 
In the nuclear sector the best limits are obtained in the ground state Zeeman  and hyperfine atomic transitions (see e.g. review  \cite{KosRus11}). Indeed, here we do not need nuclear transitions since the energy shifts produced by the tensor operators are different for different angular momentum projections. 

The aim of this section of the  present paper is to use nuclear calculations for significant improvement of the best limits on the proton LLIV constants  $c_{ab}$ using existing experimental data \cite{Romalis}  for $^{21}_{10}$Ne (see also previous  experiments \cite{Lamoreaux,Chupp,Prestage,Cs}). We will see that the LLIV tensor has a collective nature similar to the nuclear electric quadrupole moment.


Let us start from simple estimates of the quadrupole moment $Q$ and the LLIV tensor $\delta H$ in the Schmidt (single-valence-nucleon) model. The calculations similar to the calculation of $Q$  in the textbook \cite{Landau} give the following results:
 \begin{eqnarray}
 \label{Q}
 \nonumber
 Q \equiv Q_{zz}=<3 z^2 -r^2>= - \frac{j-1/2}{j+1} <r^2>\\
 =- \frac{j-1/2}{j+1} \cdot 0.009 A^{2/3} {\rm barn},\\
\nonumber
 <\delta H> \equiv C^{(2)}_0 \gamma_w m=C^{(2)}_0\frac{<3p_z^2 - p^2>}{6m}=\\
 -\frac{C^{(2)}_0}{3} \frac{j-1/2}{j+1} <\frac{p^2}{2m}>=- C^{(2)}_0 \frac{j-1/2}{j+1} \cdot 10 \,{\rm MeV},
 \label{C}
 \end{eqnarray}
where we use the  standard notation $C^{(2)}_0=c_{xx}+c_{yy}-2c_{zz}$ for the LLIV tensor interaction coefficient in the laboratory frame with the quantisation axis along $z$,  $j$ is the total angular momentum of the valence nucleon (equal to the nuclear spin $I$). As usual, the expectation values are taken for the maximal value of the angular momentum projection, $j_z=I_z=I$. Note that the results do not depend on the valence nucleon orbital angular momentum $l$ (the expression for $\gamma_w =<\delta H>/C^{(2)}_0 m$ in Ref.
\cite{KosLan99} agrees with our result after the substitution  $l=j \pm 1/2$). The numerical estimates have been presented for the following values of the parameters. For the valence nucleon $<r^2> \approx (3/5) R ^2 =0.9 A^{2/3}$ fm$^2$, where $R= 1.2A^{1/3}$ fm  is the nuclear radius, $A$ is the number of the nucleons in the nucleus, barn=100 fm$^2$=$10^{-24}$ cm$^{2}$.
The depth of the nuclear potential $V_0 \approx$ 50 MeV  \cite{BM1}. The binding energy of the valence nucleon in medium and heavy nuclei is 5 - 10 MeV. This gives us a square-well estimate of the valence nucleon kinetic energy $<p^2/2m> \approx $ 40 MeV. The relativistic parameter $K= <p^2>/m^2$ in this model is $K$=0.08. The oscillator model gives  $<p^2/2m> = m \omega_0^2 <r^2>/2 
 \approx $ 20 MeV, $K=0.04$ (here $\omega_0 \approx 40$ MeV$/A^{1/3}$ is the nuclear oscillator frequency \cite{BM1}).
The shape of the real nuclear potential is between the square well and the oscillator potential. Therefore, we assume the average values  $<p^2/2m> = $ 30 MeV,  $K=0.06$. 

The Schmidt model is not very accurate for the description of the experimental values of the electric quadrupole moments $Q$ (and the LLIV tensors).  It predicts $Q=0$  in nuclei of the experimental interest 
$^{21}$Ne, $^{9}$Be, $^{201}$Hg, $^{173}_{70}$Yb and  $^{131}_{54}$Xe
  where the valence nucleon is neutron while the experimental values of $Q$ are actually large. 
On the other hand, the experimental value of $Q$ in $^{133}$Cs is 40 times smaller than the Schmidt model value. To solve this problem  we suggest a semi-empirical approach.   Looking on the Eqs. (\ref{Q}) and (\ref{C}) we see that the LLIV tensor and the quadrupole moment are proportional to each other. This proportionality relation actually does not require applicability of the Schmidt  model.  Indeed, we may use the following equations:
\begin{eqnarray}
 \nonumber
<0| p_a|n> = i m <0|[ H, r_a]|n>=\\
i m (E_0 -E_n) <0|r_a |n>\,\\
\nonumber
 <0| p_a p_b |0>= \sum_n <0| p_a | n><n|p_b|0> =\\
 m^2 \omega_0^2 < 0|r_a r_b|0>\, 
 \label{Posc}
 \end{eqnarray}
where we have taken into account that in the oscillator potential the operator $r_a$ and $p_a$ have matrix elements 
 to the next shell only, with $|E_n-E_0|=\omega_0 \approx 40$ MeV$/A^{1/3}$ \cite{BM1}. We may sum these relations over all nucleons in the nucleus and link the total values of the LLIV tensor and quadrupole moment, so the Schmidt single-valence-nucleon  model is not needed. To account for the many-body  corrections, one may use the experimental value of the giant dipole resonance frequency  $E_n-E_0=\omega_D \approx 79$ MeV$/A^{1/3}$ \cite{BM1,Giant}  instead of the nuclear oscillator frequency
$\omega_0 \approx 40$ MeV$/A^{1/3}$  since this resonance saturates the sum rule for the dipole transitions. 
 Use of the giant dipole resonance frequency would lead to the 4 times larger  result.  Instead we will use an intermediate frequency $ \omega_0 <{\overline \omega} < \omega_D$ giving an estimate which
coincides with the Schmidt model estimate of the ratio  $<\delta H>/Q$ (see Eqs. (\ref{Q}) and (\ref{C})):
 \begin{eqnarray}
 \nonumber
 <\delta H> \equiv C^{(2)}_0 \gamma_w m=C^{(2)}_0\frac{<3p_z^2-p^2>}{6m}\\
 =C^{(2)}_0\frac{{\overline \omega}^2 m}{6}Q
 =C^{(2)}_0 \frac{1100}{A^{2/3}}\frac{Q}{{\rm barn}}{\rm MeV}\,
 \label{QC}
 \end{eqnarray}  
This relation allows us to use the experimental values of the electric quadrupole moments presented in  the tables in Ref. \cite{Tables}  to find the proton LLIV interaction contribution to the energy shifts. The neutron contribution is of the same order of magnitude since the shapes of the proton and neutron
distributions are very close. The numerical values of $Q$ and  $<\delta H>$ are presented in the Table \ref{t:LLV}.
\begin{table*}

\label{t:LLV}
\begin{ruledtabular}
\caption{ The values of the quadrupole moments $Q$ and LLIV shifts $<\delta H>$ obtained using the  Schmidt single-valence-nucleon model and the semi-empirical model which expresses $<\delta H>$ via the experimental values of $Q$} 
\begin{tabular}{lcccccccccc}
           &             & Schmidt            & Exp.              &                       & \multicolumn{3}{c}{ $<\delta H/C^{(2)}_0>_p$ (MeV) }&  \multicolumn{3}{c}{$<\delta H/C^{(2)}_0>_n$ (MeV)} \\
Nuclei & $J^\pi$ & $Q_p$ (barn) & $Q_p$ (barn) &$Q_n$ (barn) & Schmidt & From $Q$ & Recomended & Schmidt & From $Q$ & Recomended \\
\hline
$^{133}_{55}$Cs  & $\frac{7}{2}^{+}$ & -0.15 & -0.00358 & $\sim$ 0.15 & -7 & -0.15 &  1 & 0 & 1 & 1\\
$^{85}_{37}$Rb    & $\frac{5}{2}^-$    & -0.095 & 0.277  & $\sim$ 0.1  &  -6   & 15 & 10  & 0    &   15   & 10         \\
$^{87}_{37}$Rb    & $\frac{3}{2}^-$    &  -0.067 & 0.134  &    $\sim$ 0.06 & -4   & 7.6   &  5     & 0    & 7         &    5      \\
$^{131}_{54}$Xe  & $\frac{3}{2}^+$   &   0  &     -0.114  & -0.1 &   0  & -5   &   -4   & -4    &   -5      & -4         \\
$^{201}_{80}$Hg  & $\frac{3}{2}^-$    &    0  & 0.40   &  0.4    & 0    & 13   & 10  & -4  & 13  & 10   \\
$^{21}_{10}$Ne    & $\frac{3}{2}^+$   &    0   & 0.103  &    0.116   & 0    & 0.54  & 0.54   & -4     & 0.57         &   0.57      \\
\end{tabular}

\end{ruledtabular}
\end{table*}

 {\bf Spacial tensor  LLIV interaction in deformed  nuclei: the oscillator model}. Deformed nuclei may have  very large electric quadrupole moments. Looking to Eq. (\ref{QC}), one may expect a similar enhancement for the LLIV tensor. However, there is no simple proportionality relation between the LLIV tensor and quadrupole moment in this case. Indeed, the deformed oscillator corresponding to the quadrupole deformation is described by the two frequencies: $\omega_z$ and $\omega_x$. The different dependence on these two frequencies for $Q$ and $M$ is of a crucial importance. In the rotating (frozen body) reference frame
\begin{eqnarray} \label{Q0}
Q_0 &=& \sum_k<2 z_k^2 -x_k^2-y_k^2>=
 \sum_k (  \frac{2 \epsilon_z^k}{m \omega_z^2 } -  \frac{\epsilon_x^k + \epsilon_y^k}{m \omega_x^2 })\nonumber\\
&=& \frac{\hbar}{m}  \sum_k (  \frac{2 n_z^k +1}{\omega_z} -  \frac{N^k - n_z^k  +1}{\omega_x}),\\
 M_0 &=&\sum_k<2 p_{z,k}^2 -p_{x,k}^2-p_{y,k}^2>=m
 \sum_k ( 2 \epsilon_z^k - \epsilon_x^k - \epsilon_y^k)\nonumber\\
 & =&\frac{\hbar}{m}  \sum_k [ (2 n_z^k +1)\omega_z - (N^k - n_z^k  +1)\omega_x],
 \label{M0}
 \end{eqnarray} 
 where the summation goes over occupied nuclear orbitals $k$, $\epsilon_z=\hbar \omega_z (n_z+1/2)$ and $N=n_x+n_y+n_z$. The transition to the laboratory frame is given by
 \begin{equation}\label{T}
\nonumber 
Q=Q_0 \frac{(2 I-1)I }{(2I+3)(I+1)},\,M=M_0 \frac{(2 I-1)I }{(2I+3)(I+1)}.
\end{equation}
  Let us start from a simple Fermi-gas estimate assuming summation over all orbitals with $ \epsilon^k= \epsilon_x^k + \epsilon_y^k + \epsilon_z^k \le \epsilon_F$ where 
 $ \epsilon_F$ is the Fermi energy.  In this model we have $<\epsilon_x>=<\epsilon_y>=<\epsilon_z>=<\epsilon>/3$ and $Q$ is strongly enhanced ($\sim Z \delta $ times, where $\delta$ is the deformation parameter):
 \begin{equation}\label{QFermi}
Q_0\approx. \frac{2 Z <r^2>{\overline \omega}^2}{3}(\frac{1}{\omega_z^2} - \frac{1}{\omega_x^2}),
\end{equation}
where  $<r^2> =\sum_k<z_k^2 +x_k^2+y_k^2>/Z\approx (3/5) R ^2 =0.9 A^{2/3}$ fm$^2$ and  $R= 1.2A^{1/3}$ fm  is the nuclear radius. However, in the approximation 
 $<\epsilon_x>=<\epsilon_y>=<\epsilon_z>=<\epsilon>/3$ we obtain $M=0$.

 To obtain a non-zero $M$ we should do explicit summation over occupied orbitals in a specific nucleus. Quantum numbers $N$ and $n_z$ and energy ordering of the deformed oscillator orbitals including the spin-orbit interaction have been presented  in the book \cite{BM}. The results for $M_p$ and $M_n$ may still be found using the experimental values of the electric quadrupole moments.   For   $^{21}_{10}$Ne which has spin $I=3/2$, 10 protons and 11 neutrons, Eqs. (\ref{Q0},\ref{M0},\ref{T}) give the deformed oscillator value of the electric quadrupole which may be equated to the experimental value $Q=$0.103(8) barn: 
\begin{equation} \label{QNe} 
Q = \frac{\hbar}{5m} (  \frac{22}{\omega_z} -  \frac{14}{\omega_x})=0.103 \,{\rm barn}.
\end{equation}
Combined this equation with the equation $\omega_z + 2\omega_x =3 {\overline \omega} =13.03$ MeV we find $ {\overline \omega}/\omega_z=1.304$,  
${\overline \omega}/\omega_x=0.896$,
\begin{equation} \label{MpNe} 
M_p/m = \frac{\hbar}{5} ( 22\omega_z - 14\omega_x)=3.2{\rm MeV},
\end{equation}
\begin{equation} \label{MnNe} 
M_n/m = \frac{\hbar}{5} ( 25\omega_z - 16\omega_x)=3.4{\rm MeV}.
\end{equation}
Here we have used the improved formula for  the oscillator frequency ${\overline \omega} = 45 A^{-1/3} -25 A^{-2/3}$ MeV suitable for light nuclei (${\overline \omega} = $13.03 MeV   in $^{21}_{10}$Ne) from  the preprint \cite{Brown} which also presents  results of the $M_p$ and $M_n$ calculations in $^{21}_{10}$Ne using the spherical oscillator $s-d$ shell model  and the Hartree-Fock-Bogoliubov methods.

Using Eqs. (\ref{MpNe},\ref{MnNe})  we obtain the  LLIV shift in  $^{21}_{10}$Ne
$ <\delta H>=(0.54 C^{(2)}_{0,p} + 0.57 C^{(2)}_{0,n}) \, {\rm MeV}$.
 Comparing this result with an estimate $<\delta H>=-C^{(2)}_{0,n} \cdot$ 0.65 MeV used in 
the experimental paper \cite{Romalis} we  conclude that the experimental limits on the the tensor LLIV constants obtained in Ref. \cite{Romalis}, 
\begin{eqnarray}
c_X=c_{YZ}+c_{ZY}=(4.8 \pm 4.4) \cdot 10^{-29},\\ 
c_Y=c_{XZ}+c_{ZX}=-(2.8 \pm 3.4) \cdot 10^{-29},\\ 
c_Z=c_{XY}+c_{YX}=-(1.2 \pm 1.4) \cdot 10^{-29},\\
c_-=c_{XX}-c_{YY}=(1.4 \pm 1.7) \cdot 10^{-29},
\label{NeC}
 \end{eqnarray}
actually  include the linear combination   $c= - (0.83 c_p + 0.88 c_n)$ instead of $c_n$, i.e. they contain the proton LLIV constants.
This means that the limits on the proton constants are improved by  4  orders of magnitude in comparison with the previous best limits  obtained using the Cs fountain in Ref.\cite{Cs}.
Using the method described in the present paper we have calculated LLIV tensors for all deformed nuclei of experimental interest. The results will be presented in a future publication \cite{Lackenby2016}. 

In Table \ref{t:LLV} we also presented estimates for the quadruple moments of the neutron distribution (NQM).
Within the Standard model the neutron weak charge ($q_w^n=-1$) significantly exceeds the proton weak charge ($q_w^p =0.08$). Therefore, NQM may be measured in a specially designed experiment measuring parity non-conservation (PNC) in molecules or atoms. NQM generates the tensor weak interaction $W_T=W_{ik}I_iI_k$ which can mix atomic electrons states of opposite parity and angular momentum difference $J_1-J_2=2$, or molecular $\Omega$-doublet states (linear combinations of the states with the projections of $J$ on molecular axis $\Omega$=1 and    $\Omega$=-1) \cite{SF1978}. The energy interval between the $\Omega$-doublet states is extremely small, therefore, corresponding PNC effects will be strongly enhanced. The list of suitable molecules includes TaN, ThO, ThF+, HfF+, PbO, WC  and many other molecules in  $|\Omega|$=1 electron state where the heavy nucleus isotope should have nuclear spin $I>1/2$. In Ref. \cite{Flambaum2014} we proposed these molecules to measure the time-reversal invariance violating nuclear magnetic quadrupole moments  and study CP-violating interactions \cite{Flambaum2014}. Experiment with $^{232}$ThO  ($I$=0) \cite{ThO} has  allowed to improve the limit on the electron electric dipole moment by an order of magnitude. Experiments with ThF+, HfF+, PbO, WC are in progress.

To summarise, in this work we performed calculations which will give one a new interpretation of existing and future experiments searching for anisotropy in the maximal attainable speed for massive particles.
New interpretation of the experiment with $^{21}$Ne \cite{Romalis} has already allowed us to improve the limits on corresponding LLIV parameters for protons by 4 orders of magnitude.
We identified enhanced effects and proposed new experiments which  should lead to a significant increase in the accuracy of the search for violation of the Einstein Equivalence Principle  and first measurements of the quadrupole moments of the neutron distributions.

This work is  supported in part by the Australian Research Council and Gutenberg Fellowship. I am grateful to A. Brown for sending me preliminary results of their unpublished work  before it appeared in arXiv \cite{Brown} (in response to the first version of the present paper arxiv:1603.05753v1).

\end{document}